\documentclass[aps,twocolumn]{revtex4}
\usepackage{amsmath,amssymb}

\usepackage{tabularx}

\usepackage{epsfig}
\usepackage{graphicx}

\newcommand{\be}{\begin{equation}}
\newcommand{\ee}{\end{equation}}
\newcommand{\bea}{\begin{eqnarray}}
\newcommand{\eea}{\end{eqnarray}}
\newcommand{\nn}{\nonumber}

\begin{document}

\title{Brane-bulk energy exchange and the Universe as a global attractor
\footnote{This talk is based on work with G. Kofinas and Gr.
Panotopoulos \cite{kpt}}.}

\author{Theodore N. Tomaras$^{1}$\footnote{tomaras@physics.uoc.gr}}

\date{\today}

\address{$^{1}$Department of Physics and Institute of Plasma Physics, University of Crete, 71003 Heraklion, Greece}

\begin{abstract}

The assumption that our Universe is close to a late time fixed
point of the equations of cosmology, leads to a modification of
the latter to include energy exchange between the matter and the
"dark energy". The brane-world scenario provides a natural set-up
for such energy exchange and is analyzed in detail. The role of
brane-bulk energy exchange and of an induced gravity term on a
single braneworld of negative tension and vanishing effective
cosmological constant is studied. It is shown that for the
physically interesting cases of dust and radiation a unique global
attractor which can realize our present universe (accelerating and
$0\!<\!\Omega_{m0}\!<\!1$) exists for a wide range of the
parameters of the model. For $\Omega_{m0}\!=\!0.3$, independently
of the other parameters, the model predicts that the equation of
state for the dark energy today is $w_{DE,0}\!=\!-1.4$, while
$\Omega_{m0}\!=\!0.03$ leads to $w_{DE,0}\!=\!-1.03$. In addition,
during its evolution, $w_{DE}$ crosses the $w_{DE}\!=\!-1$ line to
smaller values.

\end{abstract}

\maketitle

\section{Motivation - Introduction - Conclusions}

The use of renormalization group and fixed point ideas in various
efforts to understand the symmetries or the parameter values
chosen by Nature, is not new \cite{ait}. Imagine that the
equations of cosmology have a late time fixed point. This means
that as long as the Universe is old enough, the values of its
observable quantities, simple expressions in terms of the
fundamental parameters, should be close to their values at the
fixed point. Then, the answer to the coincidence question ``why it
is that today $\Omega_{m0}$ and $\Omega_{DE,0}$ are of the same
order of magnitude'', instead of relying on a fine-tuning of
\textit{initial conditions}, it reduces to an appropriate choice
of the {\it parameters} of the theory, which should not require
any serious fine-tuning.

However, as we will now demonstrate, the equations of standard
cosmology do not accommodate our Universe as a late-time fixed
point. Specifically, I will show that
\par
{\it If the energy density of a perfect fluid with equation of
state $w\!>\!-1/3$ of any cosmological system is conserved, all
fixed points of the system with $\Omega_{m}\!\neq\! 0$ are
decelerating.} This contradicts the acceleration of our Universe.

Indeed, with $\rho$ the energy density of the perfect fluid, the
equations of cosmology are
\bea
H^{2}- 2\gamma(\rho\!+\!\rho_{DE})&=&0 \nonumber \\
\dot{\rho}+3(1+w)H\rho\!&=&\!0 \nonumber \\
\dot{\rho}_{DE}+3(1\!+\!w_{DE})H\rho_{DE}\!&=&\!0
\label{cosmology} \eea where $\gamma\!=\!4\pi G_{N}/3$. The first
is the standard Hubble equation. The second is the matter energy
conservation. The equation for $\rho_{DE}$ can always be brought
into the above form, where $w_{DE}$ is in general time as well as
model dependent. Using (\ref{cosmology}), it is straightforward to
derive $d(\Omega_{m}/\Omega_{DE})/d\ln
a\!=\!3(\Omega_{m}/\Omega_{DE})(w_{DE}\!-\!w)$ and
$2q\!=\!1\!+\!3(w\Omega_{m}\!+\!w_{DE}\Omega_{DE})$, where
$\Omega_{m}\!=\!2\gamma \rho/H^{2}$, $\Omega_{DE}\!=\!2\gamma
\rho_{DE}/H^{2}$ and $q\!=\!-\ddot{a}/aH^{2}$. At the fixed point
(denoted by $\ast$) $d(\Omega_{m}/\Omega_{DE})/d\ln a\!=\!0$. For
$\Omega_{m\ast}\!\neq\! 0$ one obtains $w_{DE\ast}\!=\!w$, and
consequently, $2q_{\ast}\!=\!1\!+\!3w\!>\!0$ Q.E.D.

Thus, independently of the cosmological model, the only way our
{\it accelerating} universe with $\Omega_{m\ast}\!\neq \!0$ can be
close to a late time fixed point is by {\it violating the standard
conservation equation of matter}.

Braneworlds are a natural set-up for such matter non-conservation.
The observable Universe is represented by a 3-brane embedded in a
higher dimensional bulk and the above violation may be the result
of energy exchange between the brane and the bulk. In particular
in five dimensions, a universe with fixed points characterized by
$\Omega_{m\ast}\!\neq\! 0$, $q_{\ast}\!<\!0$ was realized in
\cite{kkttz} in the context of the Randall-Sundrum braneworld
scenario with energy influx from the bulk. However, those fixed
points cannot represent the present universe, since they have
$\Omega_{m\ast}\!>\!2$. Here I will present a brane-bulk energy
exchange model with induced gravity, whose global attractor {\it
can} represent today's universe. Four-dimensional scenaria with
accelerating late time cosmological phase may be found in
\cite{amendola}.

{\it An interesting result}: Let us, then, generalize the
equations of cosmology to allow for violation of matter energy
conservation, due to some hypothetical "interaction". \bea
H^{2} - 2\gamma(\rho\!+\!\rho_{DE})&=&0 \nonumber \\
\dot{\rho}+3(1+w)H\rho\!&=&\!-T \nonumber \\
\dot{\rho}_{DE}+3(1\!+\!w_{DE})H\rho_{DE}\!&=&\!T \label{modified}
\eea Given the second equation of (\ref{modified}), the equation
governing $\rho_{DE}$ can always be brought into the above form,
where $w_{DE}$ is time and model dependent. If, in addition, at
the fixed point \be
H_{\ast}T_{\ast}\!\neq\!0\,,\;\;\,\dot{\rho}=\dot{\rho}_{DE}=0,\\
\label{fro}\\
\ee one obtains
\be
w_{DE\ast}=-1-\frac{1+w}{\Omega_{m\ast}^{-1}-1}.\\
\label{trikala}\\
\ee
A few comments are in order here. (a) Equation (\ref{trikala})
is {\it universal}. It does not depend on the form of $T$ or the
function $w_{DE}(t)$. (b) For $\Omega_{m\ast}\!<\!1$ equation
(\ref{trikala}) gives $w_{DE\ast}\!<\!-1$. (c) In the
phenomenologically interesting case of $w\!=\!0$ and
$\Omega_{m\ast}\!=\!\Omega_{CDM}\!=\!0.3$, i.e. if we take the
present value of matter density to include the cold dark matter,
one obtains $w_{DE\ast}\!=\!-1.4$. Finally, (d) assuming that the
dark matter is not part of $\rho_{matter}$,
$\Omega_{m\ast}\!=\!\Omega_{bar}= 0.03$, and one obtains
$w_{DE\ast}\!=\!-1.03$.
It is not surprising that once you allow for influx from the bulk,
you may obtain constant acceleration in the observable Universe.
After all, this is similar to the steady state universe. What is
surprising is that leads to the specific universal prediction for
the dark energy equation of state, which in addition, has been
favored by some recent analyses of the cosmological data.

A braneworld model with a 3-brane embedded in a 5-dimensional bulk
will be presented, whose equations will be explicitly brought into
the form of equations (\ref{modified}). For a wide range of
parameters they have a late time global attractor, representing an
accelerating universe with constant energy density on the brane,
sustained by energy influx from the bulk. As we argued above, the
fixed point satisfies (\ref{fro}), (\ref{trikala}). Moreover, via
a numerical integration it will be shown that for a wide range of
initial conditions the universe, during its evolution, crosses the
$w_{DE}\!=\!-1$ line from higher values. Interestingly enough,
these features are favored by several model-independent
\cite{staro} as well as model-dependent \cite{laura, alam,
leandros, phantom} analyzes of the astronomical data.

\section{The model}

We shall assume that we live on a 3-brane embedded in five
dimensional AdS space-time. We consider the model described by the
gravitational brane-bulk action \cite{Dvali}
\begin{equation}
S=\int \!d^{5}x\sqrt{-g}\,(M^{3}R-\Lambda)\,+\int
\!d^{4}x\sqrt{-h}\,(m^{2}\hat{R}-V),
\end{equation}
where $R,\hat{R}$ are the curvature scalars of the bulk metric
$g_{AB}$ and the induced metric $h_{AB}\!=\!g_{AB}\!-\!n_{A}n_{B}$
respectively ($n^{A}$ is the unit vector normal to the brane and
$A,B\!=\!0,1,2,3,5$). The bulk cosmological constant is
$\Lambda/2M^{3}\!<\!0$, the brane tension is $V$, and the
induced-gravity crossover scale is $r_{c}\!=\!m^{2}/M^{3}$.
\par
We assume the cosmological bulk ansatz
\begin{equation}
ds^{2}=-n(t,y)^2dt^{2}+a(t,y)^{2}\gamma_{ij}dx^{i}dx^{j}+b(t,y)^{2}dy^{2},
\end{equation}
where $\gamma_{ij}$ is a maximally symmetric 3-dimensional metric,
parameterized by the spatial curvature $k\!=\!-1,0,1$. The
non-zero components of the five-dimensional Einstein tensor are
\begin{eqnarray}
&&\!\!\!\!\!\!G_{00}=3\Big\{\frac{\dot{a}}{a}\Big(\frac{\dot{a}}{a}\!+\!\frac{\dot{b}}{b}\Big)-
\frac{n^{2}}{b^{2}}\Big[\frac{a''}{a}\!+\!\frac{a'}{a}\Big(\frac{a'}{a}\!-\!\frac{b'}{b}\Big)\Big]+
\frac{kn^{2}}{a^{2}}\!\Big\} \!\label{eq:11}\\
&&\!\!\!\!\!\!G_{ij}\!=\!\frac{a^{2}}{b^{2}}\gamma_{ij}\Big\{\frac{a'}{a}\Big(\!\frac{a'}{a}\!
+\!\frac{2n'}{n}\!\Big)-\frac{b'}{b}\Big(\!\frac{n'}{n}\!+\!\frac{2a'}{a}\!\Big)+\frac{2a''}{a}\!+\!\frac{n''}{n}\Big\} \nn \\
&&+\frac{a^{2}}{n^{2}}\gamma_{ij}\Big\{\frac{\dot{a}}{a}\Big(\frac{2\dot{n}}{n}\!-\!
\frac{\dot{a}}{a}\Big)\!-\!\frac{2\ddot{a}}{a}\!+\!\frac{\dot{b}}{b}\Big(\frac{\dot{n}}{n}\!-\!
\frac{2\dot{a}}{a}\Big)\!-\!\frac{\ddot{b}}{b}\Big\}\!-\!k\gamma_{ij} \label{eq:12} \\
&&\!\!\!\!\!G_{05}\!=\!3\Big(\frac{n'}{n}\frac{\dot{a}}{a}+\frac{a'}{a}\frac{\dot{b}}{b}-\frac{\dot{a}'}{a}\Big)
\label{eq:13}\\
&&\!\!\!\!\!G_{55}\!=\!3\Big\{\frac{a'}{a}\Big(\!\frac{a'}{a}\!+\!\frac{n'}{n}\!\Big)-\frac{b^{2}}{n^{2}}
\Big[\frac{\ddot{a}}{a}+\frac{\dot{a}}{a}\Big(\!\frac{\dot{a}}{a}\!-\!\frac{\dot{n}}{n}\!\Big)\Big]-
\frac{kb^{2}}{a^{2}}\Big\},\! \label{eq:14}
\end{eqnarray}
where primes indicate derivatives with respect to $y$, and dots
derivatives with respect to $t$. The five-dimensional Einstein
equations take the usual form
\begin{equation}
G_{AC}=\frac{1}{2M^{3}}T_{AC}|_{tot},\label{Einstein}
\end{equation}
where \begin{eqnarray} T_{C}^{A}|_{tot}\!=
T_{C}^{A}\!|_{v,B}+T_{C}^{A}\!|_{m,B}+T_{C}^{A}\!|_{v,b}\!+T_{C}^{A}\!|_{m,b}\!+T_{C}^{A}\!|_{ind}
\end{eqnarray}
is the total energy-momentum tensor,
\begin{eqnarray}
&&\!\!\!\!\!\!\!\!\!\!\!\!T_{C}^{A}|_{v,B}=\textrm{diag}(-\Lambda,-\Lambda,-\Lambda,-\Lambda,-\Lambda)\\
&&\!\!\!\!\!\!\!\!\!\!\!\!T_{C}^{A}|_{v,b}=\textrm{diag}(-V,-V,-V,-V,0)\frac{\delta(y)}{b}\\
&&\!\!\!\!\!\!\!\!\!\!\!\!T_{C}^{A}|_{m,b}=\textrm{diag}(-\rho,p,p,p,0)\frac{\delta(y)}{b}.
\end{eqnarray}
$T_{C}^{A}|_{m,B}$ is any possible additional energy-momentum in the
bulk, the brane matter content $T_{C}^{A}|_{m,b}$ consists of a
perfect fluid with energy density $\rho$ and pressure $p$, while the
contributions arising from the scalar curvature of the brane are
given by
\begin{eqnarray}
&&\!\!\!\!\!\!\!\!T_{0}^{0}|_{ind}=\frac{6m^{2}}{n^{2}}\Big(\frac{\dot{a}^{2}}{a^{2}}+\frac{kn^{2}}{a^{2}}\Big)
\frac{\delta(y)}{b}\\
&&\!\!\!\!\!\!\!\!T_{j}^{i}|_{ind}=\frac{2m^{2}}{n^{2}}\Big(\frac{\dot{a}^{2}}{a^{2}}-\frac{2\dot{a}\dot{n}}{an}+
\frac{2\ddot{a}}{a}+\frac{kn^{2}}{a^{2}}\Big)\delta_{j}^{i}\frac{\delta(y)}{b}.
\end{eqnarray}
\par
Assuming a $\mathbb{Z}_{2}$ symmetry around the brane, the singular
part of equations (\ref{Einstein}) gives the matching conditions
\begin{equation}
\!\!\!\!\!\!\!\!\!\!\!\!\!\!\!\!\!\!\frac{a_{o^{+}}'}{a_{o}b_{o}}=-\frac{\rho\!+\!V}{12M^{3}}+\frac{r_{c}}{2n_{o}^{2}}
\Big(\frac{{\dot{a}_{o}}^{2}}{a_{o}^{2}}\!+\!\frac{kn_{o}^{2}}{a_{o}^{2}}\Big)\label{eq:15}
\end{equation}
\begin{equation}
\frac{n_{o^{+}}'}{n_{o}b_{o}}\!=\!\frac{2\rho\!+\!3p\!-\!V}{12M^{3}}
+\frac{r_{c}}
{2n_{o}^{2}}\Big(\frac{2{\ddot{a}_{o}}}{a_{o}}-\frac{{\dot{a}_{o}}^{2}}{a_{o}^{2}}-
\frac{2{\dot{a}_{o}}{\dot{n}_{o}}}
{a_{o}n_{o}}-\frac{kn_{o}^{2}}{a_{o}^{2}}\Big)\label{eq:16}
\end{equation}
(the subscript o denotes the value on the brane), while from the 05,
55 components of equations (\ref{Einstein}) we obtain
\begin{equation} \label{eq:17}
\frac{n'_{o}}{n_{o}}\frac{\dot{a}_{o}}{a_{o}}+\frac{a'_{o}}{a_{o}}\frac{\dot{b}_{o}}{b_{o}}-
\frac{\dot{a}'_{o}}{a_{o}}=\frac{T_{05}}{6M^{3}}
\end{equation}
\begin{equation} \label{eq:18}
\frac{a'_{o}}{a_{o}}\Big(\!\frac{a'_{o}}{a_{o}}+\frac{n'_{o}}{n_{o}}\!\Big)-\frac{b_{o}^{2}}{n_{o}^{2}}
\Big[\frac{\ddot{a}_{o}}{a_{o}}
+\frac{\dot{a}_{o}}{a_{o}}\Big(\!\frac{\dot{a}_{o}}{a_{o}}\!-\!\frac{\dot{n}_{o}}{n_{o}}\!\Big)
\Big]-\frac{kb_{o}^{2}}{a_{o}^{2}}\!=\!\frac{T_{55}\!-\!\Lambda
b_{o}^{2}}{6M^{3}}\!,
\end{equation}
where $T_{05},T_{55}$ are the $05$ and $55$ components of
$T_{AC}|_{m,B}$ evaluated on the brane. Substituting the expressions
(\ref{eq:15}), (\ref{eq:16}) in equations (\ref{eq:17}),
(\ref{eq:18}), we obtain the semi-conservation law and the
Raychaudhuri equation
\begin{equation}
\dot{\rho}+3\frac{\dot{a}_{o}}{a_{o}}(\rho+p)=-\frac{2n_{o}^{2}}{b_{o}}T_{5}^{0}
\label{eq:22}
\end{equation}
\begin{eqnarray}
&&\!\!\!\!\!\!\!\!\!\Big(\!H_{o}^{2}\!+\!\frac{k}{a_{o}^{2}}\!\Big)
\Big[1\!-\!\frac{r_{c}^{2}(\rho\!+\!3p\!-\!2V)}{24m^{2}}\Big]
\!+\!\frac{r_{c}^{2}(\rho\!+\!3p\!-\!2V)(\rho\!+\!V)}{144m^{4}}\nn\\
&&\!\!\!\!\!\!\!\!\!+\Big(\!\frac{\dot{H}_{o}}{n_{o}}\!+\!H_{o}^{2}\!\Big)\Big[1\!-\!\frac{r_{c}^{2}}{2}
\Big(\!H_{o}^{2}\!+\!\frac{k}{a_{o}^{2}}\!\Big)\!+\!\frac{r_{c}^{2}(\rho\!+\!V)}{12m^{2}}\Big]
\!=\!\frac{\Lambda\!-\!T_{5}^{5}}{6M^{3}}, \label{eq:19}
\end{eqnarray}
where $H_{o}\!=\!\dot{a}_{o}/a_{o}n_{o}$ is the Hubble parameter of
the brane. One can easily check that in the limit $m\! \rightarrow\!
0$, equation (\ref{eq:19}) reduces to the corresponding second order
equation of the model without $\hat{R}$ \cite{kkttz}. Energy exchange between the brane and the bulk has also
been investigated in \cite{hall, hebecker, tetra}.
\par
Since only the 55 component of $T_{AC}|_{m,B}$ enters equation
(\ref{eq:19}), one can derive a cosmological system that is largely
independent of the bulk dynamics, if at the position of the brane
the contribution of this component relative to the bulk vacuum
energy is much less important than the brane matter relative to the
brane vacuum energy, or schematically
\begin{equation}
\Big|\frac{T^{5}_{5}}{\Lambda}\Big| \ll \Big|\frac{\rho}{V}\Big|.
\label{relation}\end{equation} Then, for $|\Lambda|$ not much larger
than the Randall-Sundrum value $V^{2}/12M^{3}$, the term $T^{5}_{5}$
in equation (\ref{eq:19}) can be ignored. Alternatively, the term
$T^{5}_{5}$ can be ignored in equation (\ref{eq:19}) if simply
\begin{equation}
\Big|\frac{T^{5}_{5}}{\Lambda}\Big| \ll 1.
\label{relation1}\end{equation} Note that relations (\ref{relation})
and (\ref{relation1}) are only boundary conditions for $T_{5}^{5}$,
which in a realistic description in terms of bulk fields will be
translated into boundary conditions on these fields. In the special
case where (\ref{relation}), (\ref{relation1}) are valid throughout
the bulk, the latter remains unperturbed by the exchange of energy
with the brane.
\par
One can now check that a first integral of equation (\ref{eq:19}) is
\begin{eqnarray}
&&\!\!\!\!\!\!\!\!\!\!\!\!\!\!\!
H_{o}^{4}-\frac{2H_{o}^{2}}{3}\Big(\frac{\rho\!+\!V}{2m^{2}}\!+\!\frac{6}{r_{c}^{2}}\!-\!\frac{3k}{a_{o}^{2}}\Big)
+\Big(\!\frac{\rho\!+\!V}{6m^{2}}\!-\!\frac{k}{a_{o}^{2}}\!\Big)^{2}\!+ \nn \\
&&\,\,\,\,\,\,\,\,\,\,\,\,\,\,\,\,\,\,\,\,\,\,\,\,\,\,\,\,\,\,\,\,\,\,\,\,
+\frac{4}{r_{c}^{2}}\Big(\!\frac{\Lambda}{12M^{3}}\!-\!\frac{k}{a_{o}^{2}}\!\Big)-\frac{\chi}{3r_{c}^{2}}=0,
\label{eq:20}
\end{eqnarray}
with $\chi$ satisfying
\begin{equation}
\dot{\chi}+4n_{o}H_{o}\chi=\frac{r_{c}^{2}n_{o}^{2}\,T}{m^{2}b_{o}}\Big(\!H_{o}^{2}
\!-\!\frac{\rho\!+\!V}{6m^{2}}\!+\!\frac{k}{a_{o}^{2}}\!\Big),
\label{eq:21}\end{equation} and $T\!=\!2T_{5}^{0}$ is the
discontinuity across the brane of the 05 component of the bulk
energy-momentum tensor. The solution of (\ref{eq:20}) for $H_{o}$ is
\begin{equation} \label{eq:23}
H_{o}^{2}=\frac{\rho\!+\!V}{6m^{2}}\!+\!\frac{2}{r_{c}^{2}}\!-\!\frac{k}{a_{o}^{2}}\pm
\frac{1}{\sqrt{3}r_{c}}
\Big[\frac{2(\rho\!+\!V)}{m^{2}}\!+\!\frac{12}{r_{c}^{2}}\!-\!\frac{\Lambda}{M^{3}}\!+\!\chi\Big]^{\!\frac{1}{2}}\!,
\end{equation}
and equation (\ref{eq:21}) becomes
\begin{equation} \label{eq:24}
\dot{\chi}+4n_{o}H_{o}\chi\!=\!\frac{2n_{o}^{2}\,T}{m^{2}b_{o}}\!\Big\{\!1\pm
\frac{r_{c}}{2\sqrt{3}}
\Big[\frac{2(\rho\!+\!V)}{m^{2}}\!+\!\frac{12}{r_{c}^{2}}\!-\!\frac{\Lambda}{M^{3}}\!+\!\chi\Big]^{\!\frac{1}{2}}
\!\Big\}\!.
\end{equation}
\par
At this point we find it convenient to employ a coordinate frame
in which $b_{o}\!=\!n_{o}\!=\!1$ in the above equations. This can
be achieved by using Gauss normal coordinates with $b(t,z)\!=\!1$,
and by going to the temporal gauge on the brane with
$n_{o}\!=\!1$. It is also convenient to define the parameters
\begin{eqnarray}
\lambda & = & \frac{2V}{m^{2}}+\frac{12}{r_{c}^{2}}-\frac{\Lambda}{M^{3}} \\
\mu & = & \frac{V}{6m^{2}}+\frac{2}{r_{c}^{2}} \\
\gamma & = & \frac{1}{12m^{2}} \\
\beta & = & \frac{1}{\sqrt{3}r_{c}}.
\end{eqnarray}
For a perfect fluid on the brane with equation of state $p=w\rho$
the cosmology on the brane is described by equations
(\ref{eq:22}), (\ref{eq:23}), (\ref{eq:24}), which simplify to (we
omit the subscript o in the following)
\begin{eqnarray}
&&\,\,\,\,\,\,\,\,\,\,\,\,\dot{\rho}+3(1+w)H\rho=-T \label{eq:27}\\
&&\!\!\!\!\!\!\!\!H^{2}=\mu+2\gamma \rho \pm
\beta\sqrt{\lambda\!+\!24\gamma
\rho\!+\!\chi}-\frac{k}{a^{2}}\label{eq:25}\\
&&\!\!\!\!\!\!\!\!\!\dot{\chi}+4H\chi=24\gamma T\Big(1\pm
\frac{1}{6\beta}\sqrt{\lambda\!+\!24\gamma
\rho\!+\!\chi}\Big)\label{eq:28},
\end{eqnarray}
while the second order equation (\ref{eq:19}) for the scale factor becomes
\begin{equation}
\frac{\ddot{a}}{a}=\mu-(1\!+\!3w)\gamma \rho \pm
\beta\frac{\lambda+6(1\!-\!3w)\gamma \rho}{\sqrt{\lambda+24\gamma
\rho+\chi}}\label{eq:26}.
\end{equation}
Equivalently, setting $\psi \equiv \sqrt{\lambda+24\gamma
\rho+\chi}$, equations (\ref{eq:25}), (\ref{eq:28}), (\ref{eq:26})
take the form
\begin{eqnarray}
&&\,\,\,\,\,\,\,\,\,\,\,H^{2}=\mu+2\gamma \rho \pm \beta
\psi-\frac{k}{a^{2}}
\label{eq:34}\\
&&\!\!\!\!\!\!
\dot{\psi}+2H\Big(\!\psi-\frac{\lambda+6(1\!-\!3w)\gamma
\rho}{\psi}\!\Big)=\pm \frac{2\gamma T}{\beta}
\label{eq:35}\\
&&\!\!\!\!\!\!\!\frac{\ddot{a}}{a}=\mu-(1\!+\!3w)\gamma \rho \pm
\beta \frac{\lambda+6(1\!-\!3w)\gamma \rho}{\psi}. \label{eq:32}
\end{eqnarray}
Throughout, we will assume $T(\rho)\!=\!A\rho^{\nu}$, with
$\nu>0,\,A$ constant parameters \cite{kkttz, kiritsis}. Notice that
the system of equations (\ref{eq:27})-(\ref{eq:28}) has the
influx-outflow symmetry $T\rightarrow -T$, $H\rightarrow -H$,
$t\rightarrow -t$. For $T=0$ the system reduces to the cosmology
studied in \cite{Deffayet}.
\par

We will be referring to the upper (lower) $\pm$ solution as Branch
A (Branch B). We shall be interested in a model that reduces to
the Randall-Sundrum vacuum in the absence of matter, i.e. it has
vanishing effective cosmological constant. This is achieved for
$\mu\!=\!\mp\beta\sqrt{\lambda}$, which, given that
$m^{2}V\!+\!12M^{6}$ is negative (positive) for branches A (B), is
equivalent to the fine-tuning $\Lambda\!=\!-V^{2}/12M^{3}$. Notice
that for Branch A, $V$ is necessarily negative. Cosmologies with
negative brane tension in the induced gravity scenario have also
been discussed in \cite{yuri}.
\par
Consider the case $k=0$. The system possesses the obvious fixed
point ($\rho_{*}, H_{*}, \psi_{*})=(0, 0, \sqrt{\lambda})$. However,
for $sgn(H) T<0$ there are non-trivial
fixed points, which are found by setting $\dot{\rho}=\dot{\psi}=0$ in
equations (\ref{eq:27}), (\ref{eq:35}). For $w\leq 1/3$ these are:
\begin{eqnarray}
&&\!\!\!\!\!\!\!\!\!\!\!\!\!\frac{2T(\rho_{*})^{2}}{9(1\!+\!w)^{2}\rho_{*}^{2}}=2\mu+(1\!-\!3w)\gamma\rho_{*}\nn\\
&&\,\,\,\,\,\pm
\sqrt{9(1\!+\!w)^{2}\gamma^{2}\rho_{*}^{2}+4\beta^{2}[\lambda+6(1\!-\!3w)\gamma\rho_{*}]}
\label{fp1}\\
&&\,\,\,\,\,\,\,\,\,\,\,\,\,\,\,\,\,\,\,\,\,\,\,\,H_{*}=-\frac{T(\rho_{*})}{3(1\!+\!w)\rho_{*}}\label{fp2}\\
&&\!\!\!\!\!\!\!\!\!\psi_{*}^{2}\pm
\frac{3(1\!+\!w)}{\beta}\gamma\rho_{*}\psi_{*}-[\lambda+6(1\!-\!3w)\gamma\rho_{*}]=0\label{fp3}.
\end{eqnarray}
Equation (\ref{eq:32}) gives
\begin{eqnarray}
&&\!\!\!\!\!\!\!\!\!\!\!\!\!\!\!\!\!\!\Big(\frac{\ddot{a}}{a}\Big)_{*}\!\!=
\frac{T(\rho_{\ast})^{2}}{9(1\!+\!w)^{2}\rho_{\ast}^{2}}\,,
\label{accel}
\end{eqnarray}
which is positive, and also, it has the same form (as a function of
$\rho_{\ast}$) as in the absence of $\hat{R}$. The deceleration
parameter is found to have the value \bea q_{\ast}=-1,
\label{decel}\eea which means $\dot{H}_{\ast}\!=\!0$. Furthermore,
at this fixed point we find
\begin{eqnarray}
\Omega_{m*}\equiv\frac{2\gamma\rho_{\ast}}{H_{\ast}^{2}}=\frac{18(1\!+\!w)^{2}}{A^{2}}\gamma\rho_{\ast}^{3-2\nu}.
\label{flat1}
\end{eqnarray}
Equation (\ref{fp1}), when expressed in terms of $\Omega_{m\ast}$,
has only one root for each branch \bea
\rho_{\ast}=\frac{\beta}{2\gamma}\frac{6(1\!-\!3w)\beta\pm\sqrt{\lambda}(1\!-\!3w\!-\!4\Omega_{m\ast}^{-1})}
{(2\Omega_{m\ast}^{-1}\!+\!1\!+\!3w)(\Omega_{m\ast}^{-1}\!-\!1)}.
\label{star} \eea However, it can be seen from (\ref{star}) that
for $-1\leq w \leq 1/3$ and $\Omega_{m\ast}< 1$ the Branch B is
inconsistent with equation (\ref{fp1}). On the contrary, Branch A
with $-1\leq w \leq 1/3$ and $\Omega_{m\ast}< 1$ is consistent for
$0<6(1\!-\!3w\!)\beta\!+\!\sqrt{\lambda}(1\!-\!3w\!-\!4\Omega_{m\ast}^{-1})
<3\sqrt{4(1\!-\!3w\!)^{2}\beta^{2}\!-\!(1\!+\!w\!)^{2}\lambda}$.
Thus, since we are interested in realizing the present universe as
a fixed point, Branch B should be rejected, and from now on we
will only consider Branch A. So, we have seen until now that
{\textit{for negative brane tension, we can have a fixed point of
our model with acceleration and $0<\Omega_{m\ast}<1$}}. This
behavior is qualitatively different from the one obtained in the
context of the model presented in \cite{kkttz} (for $-1/3 \!
\leq\!w\!\leq\!1/3$), where for positive brane tension we have
$\Omega_{m\ast}>2$, while for negative brane tension the universe
necessarily exhibited deceleration; therefore, in that model the
idea that the present universe is close to a fixed point could not
be realized.

\section{Critical point analysis}

We shall restrict ourselves to the flat case $k\!=\!0$. In order to study the dynamics of the system,
it is convenient to use (dimensionless) flatness parameters such that the state space is
compact \cite{goheer}. Defining
\be \omega_{m}\!=\!\frac{2\gamma\rho}{D^{2}}
\,\,\,\,\,,\,\,\,\,\,\omega_{\psi}=\frac{\beta\psi}{D^{2}}\,\,\,\,\,,\,\,\,\,\,
Z=\frac{H}{D}\,, \label{flat} \ee where
$D\!=\!\sqrt{H^{2}\!-\!\mu}$, we obtain the equations
\begin{eqnarray}
&&\,\,\,\,\,\,\,\,\,\,\,\,\,\,\,\,\,\,\,\,\,\,\,\,\,\,\,\,\,\,\,\,\,\,\,\,\,\,\,\,\,\,\,
\omega_{m}+\omega_{\psi}=1
\label{friflat}\\
&&\!\!\!\!\!\!\!\omega_{m}'\!=\!\omega_{\!m}\!\Big[\!(1\!+\!3w)(\omega_{\!m}\!\!-\!1\!)Z\!-\!\frac{
A}{\sqrt{|\mu|}}
\Big(\!\frac{|\mu|\omega_{\!m}}{2\gamma}\!\Big)^{\!\!\nu\!-\!1}\!(1\!-\!Z^{2})^{\frac{3}{2}-\nu}\nn\\
&&\,\,\,\,\,\,\,\,\,\,\,\,\,\,\,\,\,\,\,\,\,\,\,
-2Z(1\!-\!Z^{2})\frac{1\!-\!Z^{2}\!-\!3(1\!-\!3w)\beta^{2}\mu^{-1}\omega_{m}}
{1\!-\!\omega_{m}}\!\Big] \label{gerold}\\
&&\!\!\!\!\!\!\!Z'\!=\!(1\!-\!Z^{2})\Big[(1\!-\!Z^{2})
\frac{1\!-\!Z^{2}\!-\!3(1\!-\!3w)\beta^{2}\mu^{-1}\omega_{m}}{1\!-\!\omega_{m}}-1\nn\\
&& \,\,\,\,\,\,\,\,\,\,\,\,\,\,\,\,\,\,\,\,\,\,\,\,
\,\,\,\,\,\,\,\,\,\,\,\,\,\,\,\,\,\,\,\,\,\,\,\,\,\,\,\,\,\,\,\,\,\,\,\,\,\,\,\,\,\,\,\,\,\,\,\,
\,\,\,\,\,\,\,\,\,\,\,\,\,\,\,\,\,\,\,\,-\frac{1\!+\!3w}{2}\omega_{m}\Big],
\label{italy}
\end{eqnarray}
with $'\!=\!d/d\tau\!=\!D^{-1}d/dt$. Note that $-1\leq Z\leq 1$,
while both $\omega$'s satisfy $0\leq \omega\leq 1$. The
deceleration parameter is given by \be
q\!=\!\frac{1}{Z^{2}}\!\Big[\!\frac{1\!+\!3w}{2}\omega_{m}\!-\!(1\!-\!Z^{2})
\frac{\omega_{m}\!-\!\!Z^{2}\!-\!3(1\!-\!3w)\beta^{2}\mu^{-1}\omega_{m}}{1\!-\!
\omega_{m}}\!\Big]\label{greece} \ee and
$H'=-HZ(q+1)$. The system of equations
(\ref{gerold})-(\ref{italy}) inherits from equations
(\ref{eq:27})-(\ref{eq:28}) the symmetry $A\rightarrow -A$,
$Z\rightarrow -Z$, $\tau\rightarrow -\tau$. The system written in the new variables contains only three
parameters. However, going back to the physical quantities $H$, $\rho$ one will need specific values of
two more parameters.
\par
It is obvious that the points with $|Z|=1$ have $H=\infty$.
Therefore, from (\ref{eq:34}) it arises that the infinite density
$\rho\!=\!\infty$ big bang (big crunch) singularity, when it
appears, is represented by one of the points with $Z\!=\!1$
($Z\!=\!-1$). The points with $\omega_{m}\!=\!1$, $|Z|\!\neq\!
1,0$ have $\omega_{m}'\!=\infty$, $Z'\!=\infty$ and finite $\rho$,
$H$; for $w\!\leq \!1/3$, one has in addition
$\ddot{a}/a\!=\!+\infty$, i.e. divergent 4D curvature scalar on
the brane.
\par
The system possesses, generically, the fixed point (a)
$(\omega_{m\ast},\omega_{\psi\ast},Z_{\ast})\!=\!(0,1,0)$, which
corresponds to the fixed point
$(\rho_{\ast},H_{\ast},\psi_{\ast})\!=\!(0,0,\sqrt{\lambda})$ discussed above. For $\nu\!\leq\! 3/2$
there are in addition the fixed points (b)
$(\omega_{m\ast},\omega_{\psi\ast},Z_{\ast})\!=\!(0,1,1)$ and (c)
$(\omega_{m\ast},\omega_{\psi\ast},Z_{\ast})\!=\!(0,1,-1)$. All these
critical points are either non-hyperbolic, or their characteristic
matrix is not defined at all; thus, their stability cannot be
studied by first order perturbation analysis. In cases like these, one can find non-conventional
behaviors (such as saddle-nodes and cusps \cite{perko}) of the flow-chart near the critical points.
There are two more candidate fixed points (d)
$(\omega_{m\ast},\omega_{\psi\ast},Z_{\ast})=(1,0,1)$ and (e)
$(\omega_{m\ast},\omega_{\psi\ast},Z_{\ast})=(1,0,-1)$, whose
existence cannot be confirmed directly from the dynamical system,
since they make equations (\ref{gerold}), (\ref{italy})
undetermined. Apart from the above, there are other critical points given by \bea
&&\,\,\,\,\,\,\,\,\,\,\,\,\,\,\,\,\,\,\,\,
\frac{A}{\sqrt{|\mu|}}\Big(\!\frac{|\mu|\omega_{m\ast}}{2\gamma}\!\Big)^{\!\nu-1}\!=
-\frac{3(1\!+\!w)\,Z_{\ast}}{(1\!-\!Z_{\ast}^{2}) ^{\frac{3}{2}-\nu}}\label{jack}\\
&&\!\!\!\!\!\!\!
(1\!+\!3w\!)\omega_{\!m\ast}^{2}\!\!+\!(1\!-\!3w\!)\Big[\!1\!-\!\frac{6\beta^2}{\mu}
\!(1\!-\!Z_{\ast}^{2})\!\Big]\omega_{\!m\ast}\!\!-\!2[1\!-\!(1\!-\!Z_{\ast}^{2})^{2}]\nn\\
&&\,\,\,\,\,\,\,\,\,\,\,\,\,\,\,\,\,\,\,\,\,\,\,\,\,\,\,\,\,\,\,\,\,\,\,\,\,\,\,\,\,\,\,\,\,\,\,\,\,
\,\,\,\,\,\,\,\,\,\,\,\,\,\,\,\,\,\,\,\,\,\,\,\,\,\,\,\,\,\,\,\,\,\,\,\,\,\,\,\,\,\,\,\,\,\,\,\,\,\,
\,\,\,\,\,\,\,\,\,\,\,\,\,\,\,\,\,=\!0.\label{office} \eea They exist only for $A Z_{\ast}\!<\!0$ and
correspond to the ones given by equations
(\ref{fp1})-(\ref{fp3}). For the physically interesting case $w\!=\!0$ with influx we scanned the parameter
space and were convinced that for $\nu\!\neq\!3/2$ there is always only one fixed point; for $\nu\!<\!3/2$
this is an attractor ($\textsf{A}$), while for $\nu\!>\!3/2$ this is a saddle ($\textsf{S}$).
For $w\!=\!0$, $\nu\!=\!3/2$ there is either one fixed point (attractor) or no fixed points, depending on the
parameters. For the other characteristic value $w\!=\!1/3$, we concluded that for $\nu\!<\!3/2$ there is only one
fixed point (attractor), for $\nu\!>\!2$ there is only one fixed point (saddle), while for $3/2\!<\!\nu\!<\!2$
there are either two fixed points (one attractor and one saddle) or no fixed points at all, depending on the
parameters. For $w\!=\!1/3$, $\nu\!=\!3/2$ there is either one fixed point (attractor) or no fixed points.
Finally, for $w\!=\!1/3$, $\nu\!=\!2$ there is either one fixed point (saddle) or no fixed points. These results
were obtained numerically for a wide range of parameters and are summarized in Tables 1 and 2.
\vspace{0.3cm}
\begin{center}
\begin{tabular}{|c|c|c|c|}
\hline
 & $\nu<3/2$ & $\nu=3/2$ & $\nu>3/2$  \\ \hline
 No. of F.P. & 1 & 0 or 1 & 1 \\ \hline
 Nature & \textsf{A} & \,\,\,\,\,\,\,\,\,\,\,\,\textsf{A} & \textsf{S} \\
\hline
\multicolumn{4}{l}{Table 1: The fixed points for w=0, influx}
\end{tabular}
\end{center}
\vspace{0.05cm}
\begin{center}
\begin{tabular}{|c|c|c|c|c|c|}
\hline
 & $\nu\!<\!3/2$ & $\nu\!=\!3/2$ & $3/2\!<\!\nu\!<\!2$ & $\nu\!=\!2$ & $\nu\!>\!2$  \\ \hline
 No. of F.P. & 1 & 0 or 1 & \!\!0 or 2 & 0 or 1 & 1\\ \hline
 Nature & \textsf{A} & \,\,\,\,\,\,\,\,\,\,\,\,\textsf{A} & \,\,\,\,\,\,\,\,\,\,
 \textsf{A},\textsf{S} & \,\,\,\,\,\,\,\,\,\,\,\textsf{S} & \textsf{S} \\
\hline
\multicolumn{6}{l}{\,\,\,\,\,\,\,\, Table 2: The fixed points for w=1/3, influx}
\end{tabular}
\end{center}
\vspace{0.15cm}
The approach to an attractor described by the linear approximation of (\ref{gerold})-(\ref{italy})
is exponential in $\tau$ and takes infinite time $\tau$ for the universe to reach it. Given that near
this fixed point the relation between the cosmic time $t$ and the
time $\tau$ is linear, we conclude that it also takes infinite
cosmic time to reach the attractor.
\par
Defining $\epsilon\!=\!sgn(H)$, we see from (\ref{gerold})-(\ref{italy}) that the lines $Z \!=\! \epsilon$ ($\nu\!\leq\! 3/2$),
$\omega_{m}\!=\! 0$ are orbits of the system. Furthermore, the
family of solutions with $Z\!\approx \!\epsilon$ and
$dZ/d\omega_{m}\!=\!Z'/\omega_{m}'\!\approx\! 0$ is approximately
described for $\nu\!<\!3/2$ by
$\omega_{m}'\!=\!\epsilon(1\!+\!3w)\omega_{m}(\omega_{m}\!-\!1)$, and thus, they move away from the point
$(\omega_{m\ast},Z_{\ast})\!=\!(1,1)$, while they approach the point
$(\omega_{m\ast},Z_{\ast})\!=\!(1,-1)$. In addition, the solution of this
equation is
$\omega_{m}\!=\![1\!+\!ce^{\epsilon(1\!+\!3w)\tau}]^{-1}$, with
$c\!>\!0$ an integration constant. Using this solution in
equation $H'/H\!=\!-Z (q\!+\!1)$ we find that for $w\!=\!1/3$,
$H/H_{o}\!=\!\sqrt{\omega_{m}}/(1\!-\!\omega_{m})$, where $H_{o}$
is another integration constant. Then, the equation for $\omega_{m}(t)$ becomes
$d\omega_{m}/dt\!=\!-2\epsilon\omega_{m}\sqrt{H_{o}^{2}\omega_{m}\!-\!\mu(1\!-\!\omega_{m})^{2}}$,
and can be integrated giving $t$ as a function of $\omega_{m}$ or
$H$. Therefore, in the region of the big bang/big crunch singularity
one obtains $a(t)\!\sim \!\sqrt{\epsilon t}$, $\rho(t)\!\sim\!t^{-2}$, as in the standard radiation
dominated big-bang scenario. This means that for $\nu\!<\!3/2$ the energy exchange has no observable effects
close to the big bang/big crunch singularity.
\begin{figure}[h!]
\centering
\begin{tabular}{cc}
\includegraphics*[width=240pt, height=180pt]{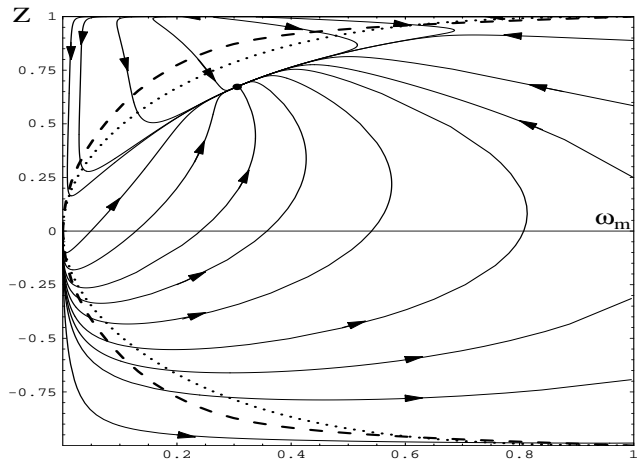}&%
\end{tabular}
 \caption{
Influx, $w\!=\!0$, $\nu\!<\!3/2$. The arrows show the
direction of increasing cosmic time. The dotted line
corresponds to $w_{DE}\!=\!-1$. The region inside (outside) the dashed line corresponds to acceleration
(deceleration). The region with $Z\!>\!0$ represents expansion, while
$Z\!<\!0$ represents collapse. The present universe is supposed to be close to the global attractor.}
\end{figure}
\begin{figure}[h!]
\centering
\begin{tabular}{cc}
\includegraphics*[width=240pt, height=180pt]{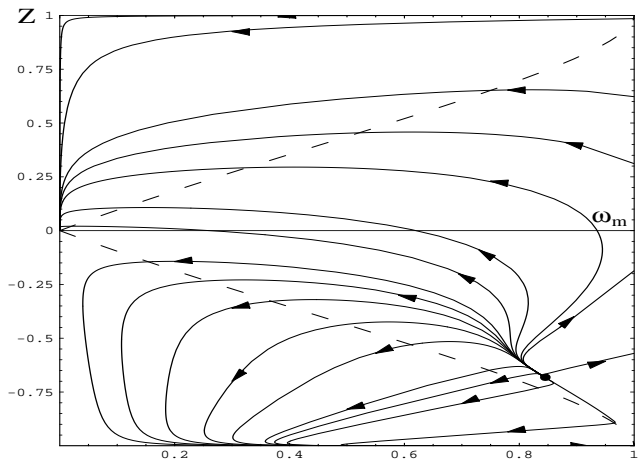}&%
\end{tabular}
 \caption{
Outflow, $w\!=\!1/3$, $\nu\!<\!3/2$. The arrows show the
direction of increasing cosmic time. The region inside (outside) the dashed line corresponds to acceleration
(deceleration). The region with $Z\!>\!0$ represents expansion, while
$Z\!<\!0$ represents collapse.}
\end{figure}
\par
Since our proposal relies on the existence of an attractor, we shall restrict ourselves to the case
$\nu\!<\!3/2$. It is convenient to discuss the four possible cases separately:
\newline
(i) $w\!=\!0$ with influx. The generic behavior of the solutions
of equations (\ref{gerold})-(\ref{italy}) is shown in Figure 1. We
see that all the expanding solutions approach the global
attractor. Furthermore, there is a class of collapsing solutions
which bounce to expanding ones. Finally, there are solutions which
collapse all during their lifetime to a state with finite $\rho$
and $H$. The physically interesting solutions are those in the
upper part of the diagram emanating from the big bang
$(\omega,Z)\!\approx\!(1,1)$. These solutions start with a period
of deceleration. The subsequent evolution depends on the value of
$3\beta^{2}\!/|\mu|$, which determines the relative position of
the dashed and dotted lines. Specifically, for
$3\beta^{2}\!/|\mu|\!>\!1$ (the case of Figure 1) one
distinguishes two possible classes of universe evolution. In the
first, the universe crosses the dashed line entering the
acceleration era still with $w_{DE}\!>\!-1$, and finally it
crosses the dotted line to $w_{DE}\!<\!-1$ approaching the
attractor. In the second, while in the deceleration era, it first
crosses the dotted line to $w_{DE}\!<\!-1$, and then the dashed
line entering the eternally accelerating era. For
$3\beta^{2}\!/|\mu|\!\leq\!1$, the dotted line lies above the
dashed line, and, consequently, only the second class of
trajectories exists. To connect with the discussion in the
introduction, notice that the Friedmann equation (\ref{eq:34}) can
be written in the form (\ref{cosmology}) with dark energy
$\rho_{DE}\!=\!(\beta\psi\!+\!\mu)/2\gamma$. Using (\ref{eq:35}),
the equation for $\rho_{DE}$ takes the form (\ref{modified}) with
\be
\!w_{DE}\!=\!\frac{-1}{3(1\!-\!\omega_{m})}\Big[\!2Z^{2}\!-\!\omega_{m}\!-\!1\!-\!6(1\!-\!3w)\frac{\beta^{2}}{\mu}
\frac{\omega_{m}(1\!-\!Z^{2})}{Z^{2}\!-\!\omega_{m}}\!\Big]\!.
\label{dark}\ee  The global attractor (\ref{fp1})-(\ref{fp3})
satisfies relations (\ref{fro}) and consequently, $w_{DE}$ evolves
to the value $w_{DE\ast}$ given by (\ref{trikala}). As for the
bouncing solutions, they approach the attractor after they cross
the line $Z^{2}\!=\!\omega_{m}$, where $w_{DE}$ jumps from
$+\infty$ to $-\infty$; however, the evolution of the observable
quantities is regular.
\newline
(ii) $w\!=\!0$ with outflow. The generic behavior in this case is obtained from Figure 1
by the substitution $Z\!\rightarrow\! -Z$ and $\tau\rightarrow -\tau$, which reflects the diagram
with respect to the $\omega_{m}$ axis and converts attractors to repelers.
\newline
(iii) $w\!=\!1/3$ with outflow. Figure 2 depicts the flow diagram of this case. Even though in the case of
radiation in general $w_{DE}\!>\!-1/3$ from equation (\ref{dark}), there are both acceleration and deceleration
regions. Furthermore, from equation (\ref{trikala}) it is
$\Omega_{m\ast}\!>\!1$.
\newline
(iv) $w\!=\!1/3$ with influx. This arises like in (ii) by reflection of Figure 2 and resembles Figure 1.

\vspace{1cm}

\section{Open questions}

There is a number of features of the present scenario, which
require further analysis. (a) One is the question of stability of
the negative tension brane. In a fixed flat background, such a
brane would be obviously unstable, by the formation of wild
ripples on the brane. Such wild fluctuations increase the area of
the brane and, with negative tension, have energy unbounded from
below. However, here the situation is different, since the brane
is in an AdS background and the effective cosmological constant on
it is zero. The stability analysis of our scenario is currently
under study. (b) It would be interesting to investigate if the
partial success of the present scenario persists after one tries
to fit the supernova data, the detailed CMB spectrum \cite{miz}
and nucleosynthesis. (c) Finally, the construction of the complete
higher dimensional theory, the specification of the nature of the
content of the bulk and of the mechanism of energy exchange with
the brane is another set of crucial open questions, which we hope
to deal with in the not too distant future.

\[ \]
{\bf Acknowlegements.} Supported in part by the EU grant
MRTN-CT-2004-512194.

\end{document}